\documentclass{JHEP3} 

\JHEPspecialurl{http://jhep.sissa.it/JOURNAL/JHEP3.tar.gz}

\usepackage{epsfig,multicol,bbm}

\def\r{\rho}
\def\a{\alpha}

\def\b{\beta}

\def\m{\mu}
\def\n{\nu}

\newcommand\fverb{\setbox\fverbbox=\hbox\bgroup\verb}
\newcommand\fverbdo{\egroup\medskip\noindent%
            \fbox{\unhbox\fverbbox}\ }
\newcommand\fverbit{\egroup\item[\fbox{\unhbox\fverbbox}]}
\newbox\fverbbox

\newcommand{\be}{\begin{eqnarray}}
\newcommand{\ee}{\end{eqnarray}}

\title{Relativistic Quantum Gravity at a Lifshitz Point}

\author{Cristiano Germani\\
    LUTH, Observatoire de Paris, CNRS UMR 8102, Universit\'e Paris Diderot, 5 Place
Jules Janssen,\\ 92195 Meudon Cedex, France\\
    E-mail: \email{cristiano.germani@obspm.fr}}
\author{Alex Kehagias\\
 Physics Division, National Technical University of Athens, \\
15780 Athens,  Greece\\
E-mail: \email{kehagias@central.ntua.gr}}
\author{Konstadinos Sfetsos\\
Department of Engineering Sciences, University of Patras\\
26110 Patras, Greece\\
E-mail: \email{sfetsos@upatras.gr}}

\abstract{We show that the Ho\v{r}ava theory for the completion of General
 Relativity at UV scales can be interpreted as
a gauge fixed theory, and it
 can be extended to an invariant theory under the full group of four-dimensional
diffeomorphisms. In this respect,  although being fully relativistic, it results to be locally
anisotropic in the time-like and space-like directions defined by a
family of irrotational observers. We show that this theory propagates generically
three degrees of freedom: two of them are related to the four-dimensional
diffeomorphism invariant graviton (the metric) and one is related to a propagating
 scalar mode.
Finally, we note that in the present formulation,
 matter can be consistently coupled to gravity.}

\begin{document}


\section{Introduction}

The main difficulty for a perturbative renormalization of General
Relativity (GR) is that
self-gravitational couplings are irrelevant operators at the Gaussian  fixed point. Although non-trivial
fixed points for the GR coupling constants might exist, corresponding to an asymptotically safe theory
 \cite{Weinberg},\cite{Nied},\cite{Rizzo},
they are still very difficult to find \cite{Percacci}.

It is instructive to recall why the Gaussian fixed point makes gravity non-renormalizable.
Consider the linearized gravity
$g_{\m\nu}=\eta_{\mu\nu}+\bar g_{\mu\nu}$,
where $\eta_{\mu\nu}$ represent the flat metric. The Einstein--Hilbert
action can then be schematically
expanded as \cite{renorm}
\be\label{action1}
S=\frac{1}{2\kappa^2} \int d^4x \left[(\partial \bar g)^2+
(\partial \bar g)^2 \bar g+\ldots \right]\ ,
\ee
where the second term is a typical self-interaction term and $\kappa$ is
the gravitational coupling constant of negative mass dimension, i.e.
$[\kappa]=-1$. Let us now use the Gaussian fixed point at UV.
For perturbative renormalization, one defines $\tilde g=\bar g/\kappa$ so that, for
energies
 up to the cut-off of the theory, the theory can be expanded in power
 of $\kappa$ around the free field theory (Gaussian fixed point).
Expanding around the fixed point we have
\be
S=\frac{1}{2} \int d^4x \left[(\partial \tilde g)^2+\kappa (\partial \tilde g)^2 \tilde g+\ldots
\right]\ .
\ee
Since $\kappa$ has dimensions of length one expects that, under
renormalization, $\kappa\propto 1/E$, at least for energies higher than the Planck mass
\cite{Percacci}. This implies that the self-coupling is an irrelevant operator that spoils
the renormalizability of the theory at the Gaussian fixed point \cite{renorm}.

The key idea for a perturbative renormalization of gravity is therefore to find
non-trivial fixed points. For example, new UV fixed points appear
whenever the Einstein--Hilbert action is modified by the introduction of
higher order four-dimensional curvatures \cite{stelle}. However,
generically, in this way one introduces ghost states
due to the higher order temporal derivatives of the metric.
Significant progress towards curing this bad behaviour was made by Ho\v{r}ava \cite{hor2}, whose
elegant idea was to write down a theory that treats
undemocratically space and time in an anisotropic way. Precursors of this idea are
Lorentz violating theories \cite{Lorentz} and, more recently, theories with
ghost condensation \cite{ghost}.

In the Ho\v{r}ava theory, one hopes to obtain
a higher spatial derivative theory of gravity, but canonical in
time, which is power counting renormalizable (due to the change in mass dimension of the gravitational
coupling)
without propagating ghosts modes.
To illustrate this let us suppose that one can find a modification of GR such that the action
(\ref{action1}) is replaced by
\be
\label{action2}
S=\frac{1}{2\kappa^2} \int d^4x \left[-\dot {\bar g}^2+\partial_i \bar g\partial^i \bar g+\alpha^{2(z-1)}
\left(\partial_{i_1}\ldots\partial_{i_z} \bar g\right)^2+(-\dot {\bar g}^2+\partial_i \bar g\partial^i
\bar g) \bar g+\ldots \right]\ ,
\ee
where we have explicitly divided the action into parts containing only
time ($\dot{}$) and space ($\partial_i$) derivatives.
The constant $z\geqslant 2$ is called the anisotropy exponent by the condensed matter community,
where these kind of theories were first employed.
The coupling constant for the higher spatial derivative part of the action has negative
mass units, i.e. $[\alpha]=-1$.

One can now show that, at the level of the Landau theory, the fixed point of (\ref{action2})
is the Lifshitz point (for an introduction of the Landau theory
at the Lifshitz point see, for example, \cite{lif}, whereas for a quite extensive discussion in the
present context see \cite{visser}).
The UV free theory related to (\ref{action2}) is
\be
S=\frac{1}{2\kappa^2} \int d^4x \left[-\dot {\bar g}^2+\alpha^{2(z-1)} \left(\partial_{i_1}\ldots
\partial_{i_z} \bar g\right)^2 \right]\ .
\ee
Following the same steps as before we will now make the redefinition
$\tilde g=\alpha^{z-1} \bar g/\kappa$ and choose units of time (time re-scaling) $\tilde
t=t\ \alpha^{z-1}$.
With these definitions our interacting theory becomes
\be
\label{action22}
S=\frac{1}{2}\int d^4x \left[-\dot {\tilde g}^2
+\lambda_1\partial_i \tilde g\partial^i \tilde g+\left(\partial_{i_1}\ldots\partial_{i_z} \tilde g\right)^2
+\lambda_2(-\dot {\tilde g}^2+\lambda_1\partial_i \tilde g\partial^i \tilde g)\tilde g+\ldots \right]\ ,
\ee
where the new physical coupling constants
$\lambda_1=\alpha^{-2(z-1)}$ and $\lambda_2=\kappa\alpha^{-(z-1)}$
have now positive mass dimensions!
The dangerous irrelevant operator of GR
has now been turned into a relevant operator at the Lifshitz point. This theory is then power counting
renormalizable.

The innovative result of Ho\v{r}ava in \cite{hor1,hor2} was to realize that a Landau model for gravity,
with critical exponent $z=3$, could be written at the price of giving up the full diffeomorphisms
 invariance of GR.
In this paper we will show that this requirement is actually not necessary and the theory may be written
in a covariant, four-dimensional diffeomorphism invariant way.

\section{Relativistic propagator at the Lifshitz point}

The properties of the action (\ref{action2}) are captured by the fact that the free theory, at
the Lifshitz point,
has schematically the following scaling for the propagator (ignoring the tensorial structure)
\be
\label{propagator}
P=-\frac{1}{\omega^2-k_i k^i-\alpha^{4}(k^i k_i)^3}\ ,
\ee
where we have already specialized to the $z=3$ case, which is particularly interesting for
 constructing a power counting renormalizable theory of gravity.
In the UV (high energies, small length scales), the propagator is determined by the following scaling
\be
P\simeq -\frac{1}{\omega^2-\alpha^4(k^i k_i)^3}\ .
\ee
The form of the propagator seems to be related to the very
profound assumption that gravity is not fully covariant under space-time diffeomorphisms.
Is that true? It depends. In the following we will show that the propagator (\ref{propagator})
can be re-written in fully diffeomorphism invariant form. The key
idea is to realize that the form of the propagator (\ref{propagator}) is related
to a specific choice of coordinates.

Let us suppose that we can construct a fully diffeomorphism invariant theory of gravity with UV
completion, having  the following propagator for the linearized graviton
on a fixed background at UV scales
\be
 \label{relativistic}
P_{\rm rel}\simeq -\frac{1}{k^2-\alpha^4(\bar h^{\alpha\beta} k_\alpha k_\beta)^3} \ .
\ee
Here we have introduced
\be
h_{\alpha\beta}=g_{\alpha\beta}+n_{\alpha} n_\beta\ ,
\ee
which is the space-like metric orthogonal to a time-like direction $n^\alpha$, i.e. $h_{\a\b}n^\b=0$
and have also used a bar notation for the background quantities.

In order to get the propagator (\ref{relativistic}), the physical fields of our gravitational theory
should be a metric $h_{\alpha\beta}$
parametrized by a ``time'' parameter and a non-dynamical form $n_\alpha$.
To achieve this, we consider a foliation of
spacetime determined by the gradient of a scalar ($\chi$) that can be later used to
define time by fixing the choice of coordinates. In this way the local rest frame orthogonal to $n^\alpha$
does not ``rotate'' along $n^\alpha$. This requirement ``preserves''
the spatial properties of the foliation.\footnote{Note that, any other
time-like vector would introduce an extra degree
of freedom changing the form of the free propagator.}
In other words, this foliation, known as slicing of
spacetime \cite{manyfaces}, defines a three-dimensional sub-Riemannian spatial manifold $\Sigma_\chi$
for each foliation parametrized by $\chi$,
whose spatial metric is the inverse of $h^{\alpha\beta}$.
Namely, in this decomposition,
the projected Christoffel symbols of
the four-dimensional metric correspond to the three-dimensional Christoffel symbols related to the spatial metric.
With this structure, one may projects four-dimensional tensors to the three-dimensional slices ($\Sigma_\chi$) and defines compatible covariant derivatives $D$ on $\Sigma_\chi$ with $h_{\alpha\beta}$ by using the projector $h^\alpha_\beta=g^{\alpha\gamma}h_{\gamma \beta}$ as follows
\be
D_\gamma {T^{\alpha_1\ldots \alpha_k}}_{\beta_1\ldots\beta_\ell}=h^{\alpha_1}_{\delta_1}\cdots
h^{\kappa_\ell}_{\beta_\ell}h^{\sigma}_{\gamma}\nabla_\sigma
{T^{\delta_1\ldots \delta_k}}_{\kappa_1\ldots\kappa_\ell}\, . \label{dnab}
\ee
The non-trivial task is now to write down an action in which $n_\alpha$ is non-dynamical
when the renormalizable field $h_{\alpha\beta}$ is employed.
Note however that, in order to have a covariant theory, $n_\alpha$ must
be dynamical (in the equation of motion) if the spacetime metric $g_{\alpha\beta}$ is instead used
\cite{wald}.\footnote{We thank Jose M. Martin-Garcia for this comment.}
Recapitulating, in order to restore a broken symmetry
(the full diffeomorphism invariance) an extra field, the vector $n^\alpha$, must be introduced into the theory. This is reminiscent of the Stuckelberg formalism to restore a $U(1)$ gauge symmetry in a massive $U(1)$ theory.

As already stressed before, the normalized form $n_\alpha$ must satisfy the conditions of zero vorticity of the space-like foliation (or equivalently the Frobenius integrability conditions \cite{gorg}). Explicitly
\be
{\cal F}_{\m\n}\equiv h^\alpha_\mu h^\beta_\nu\nabla_{[\alpha}n_{\beta]}=0\ .
\label{2-8}
\ee
The general solution of (\ref{2-8}) defines the form $n_\alpha$ we are looking for to be
\be\label{def}
n_{\alpha}=-N\partial_\alpha\chi\ ,
\ee
where the lapse $N$ encodes the normalization condition
\be
n_\alpha n^\alpha=-1\ ,
\label{2-7}
\ee
and, repeating ourselves, $\chi$ defines the foliation of the spacetime.

Note that fixing $N=f(\chi)$ overconstraints the system
and corresponds to the ``projectability'' condition of Ho\v{r}ava.

If $\chi$ is now taken as a time coordinate one can use the following adapted coordinate system (ADM
decomposition \cite{ADM}, latin indices below run from $1$ to $3$)
\be
\label{adm}
ds^2=-N^2dt^2+h_{ij}(dx^i+N^idt)(dx^j+N^jdt)\ .
\ee
We remark that the choice of coordinates has
been partially fixed by requiring that $\chi=t$.
The ADM-metric can however still transform under foliation preserving
diffeomorphisms.

In this decomposition, the linearized gravity (which is now fixed to be transverse to $n^\alpha$)
 is divided into the following three
modes: a tensorial mode ($h_{ij}$), a vector mode ($N^i$) and a scalar mode ($N$).
In GR only the tensorial mode propagates and, as pointed out, we require that our theory keeps this property.

Let us suppose that the theory we are after admits a flat background. By appropriately choosing the
spatial coordinates in the metric (\ref{adm}), one obtains the Minkowski
line-element $ds^2=-dt^2+dx^2+dy^2+dz^2$.
In this gauge, we easily find $P_{\rm rel}=P$. However,
since $P_{\rm rel}$ is now a four-dimensional
scalar, it does not change under spacetime coordinate transformations.

The original Ho\v{r}ava idea was instead to keep invariant the non-relativistic propagator $P$
by effectively ``freezing-out'' the choice of foliation. In order to do that, in \cite{hor2}, only
foliation preserving diffeomorphisms where allowed (the ADM gauge freedom) restricting the covariance of
 the gravity theory.
Here instead $P_{\rm rel}$ is a space-time scalar and, therefore, all its properties
 are invariant under
coordinate transformations.
Of course there is still a preferred choice of (coordinate) gauge to check the renormalizability
properties of the theory, namely the ADM gauge. Nevertheless, by choosing the
latter, one has to remember that the group of diffeomorphisms has been partially
used by fixing time, contrary to what happen in General Relativity. In fact in GR, where
 there is no explicit dependence upon $n$, one can always take $\chi=t$, which just fixes the
foliation. The theory we propose instead depends upon the two dynamical fields
$g_{\alpha\beta}$ and $n_{\alpha}$. In this case, the choice $\chi=t$ can be done
only at the price of partially using the diffeomorphisms group! This has an
 interesting consequence as we shall see later.

Although diffeomorphism invariant, a propagator of the form
 (\ref{relativistic}), treats anisotropically the space-like and time-like portions
 of the background manifolds (the ``inner'' and ``outer'' regions of local light-cones)
 in compatibility with causality. We would like to stress once more that as opposed
 to the non-relativistic Ho\v{r}ava formulation
(where time direction has to be kept fixed), our anisotropy of space and time foliations
is again fully relativistic (i.e. invariant under diffeomorphisms).

\section{The Ho\v{r}ava theory}

To proceed, let us consider the ADM decomposition of the metric as in (\ref{adm}),
where the dynamical fields $h_{ij},N$ and $N^i$ have
scaling mass dimensions $0,0$ and $2$, respectively.
The Ho\v{r}ava's theory \cite{hor2} with a soft violation of the detailed
balance condition (represented by the last term) is
\be
S &= & \int dt d^3 x
\sqrt{h}N\left\{\frac{2}{\kappa^2}\left(K_{ij}K^{ij}-\lambda
K^2\right)-\frac{\kappa^2}{2w^4}C_{ij}C^{ij}+\frac{\kappa^2
\mu}{2w^2}\epsilon^{ijk} {\cal R}_{i\ell} \nabla_{j}{\cal R}^{\ell}{}_k
\right.
\nonumber \\
&&\left. -\frac{\kappa^2\mu^2}{8} {\cal R}_{ij}
{\cal R}^{ij}+\frac{\kappa^2 \mu^2}{8(1-3\lambda)}
\left(\frac{1-4\lambda}{4}{\cal R}^2+\Lambda_W {\cal R}-3
\Lambda_W^2\right)+\mu^4 {\cal R}\right\}\ ,
\label{hor}
 \ee
where
\be
K_{ij}=\frac{1}{2N}\left(\dot{h}_{ij}-\nabla_i
N_j-\nabla_jN_i\right)\ ,
 \ee
is the second fundamental form (extrinsic curvature) and
\be
 C^{ij}=\epsilon^{ik\ell}\nabla_k
\left({\cal R}^{j}{}_\ell-\frac{1}{4}{\cal R} \delta^j_\ell\right)\ ,
\label{cot}
 \ee
is the Cotton tensor. Moreover,  $\kappa,\lambda,w$ are dimensionless
coupling constants, $\mu,\Lambda_W$ are dimensionfull constants of
mass dimensions $[\mu]=1,[\Lambda_W]=2$
and  ${\cal R}_{ij}$ and ${\cal R}$ are the three dimensional Ricci and scalar curvatures
 related to $h_{ij}$.
Introducing now the coordinate time $x^0=c\, t$, $c$ and the effective Newton's
constant are given by
\be
c=\frac{\kappa^2\mu}{4}\sqrt{\frac{\Lambda_W}{1-3\lambda}}\ , \qquad   G_N=\frac{\kappa^2}{32\pi c}\ .
\ee
Moreover, the action (\ref{hor}) can be split into a kinetic $S_{\rm{kin}}$ and a potential term
$S_V$ according to $S=S_{\rm{kin}}+S_V$, where
\be
S_{\rm{kin}}=\int dt d^3 x
\sqrt{h}N\left[\frac{2}{\kappa^2}\left(K_{ij}K^{ij}-\lambda
K^2\right)\right]
\ee
and
\be
S_V&=&\int dt d^3 x
\sqrt{h}N\left\{
-\frac{\kappa^2}{2w^4}C_{ij}C^{ij}+\frac{\kappa^2
\mu}{2w^2}\epsilon^{ijk} {\cal R}_{i\ell} \nabla_{j}{\cal R}^{\ell}{}_k  -\frac{\kappa^2\mu^2}{8} {\cal R}_{ij}
{\cal R}^{ij}
\right.
\nonumber \\
&&\hspace{1cm}\left.+\frac{\kappa^2 \mu^2}{8(1-3\lambda)}
\left(\frac{1-4\lambda}{4}{\cal R}^2+\Lambda_W {\cal R}-3
\Lambda_W^2\right)+\mu^4 {\cal R}\right\}\ .
\label{hor1}
 \ee
This theory is invariant under the group of restricted diffeomorphisms
\be\label{restr}
\delta x^i=\zeta^i(x,t)\ ,\qquad \delta t=f(t)\ ,
\ee
since the coordinates are kept adapted to $n^\alpha$, i.e., in this theory, the slicing form is fixed by the choice of time parametrization.

In (\ref{hor}) it seems to exists a preferred direction (time).
For these reasons the Ho\v{r}ava theory has
been said to be non-relativistic. Indeed, the theory so formulated does
not enjoy full four-dimensional covariance.
This is reminiscent of a
$U(1)$ gauge theory with action
\be
S=\int d^4 x \sqrt{-g}\left(-\frac{1}{4}F_{\mu\nu}F^{\mu\nu}+\frac{1}{2a}( \partial_\mu A^\mu)^2\right)\ ,
\ee
which is invariant under the "restricted" $U(1)$ symmetry
$A_\mu\to A_\mu+\partial_\mu\theta$, where the gauge parameter $\theta $ satisfies $\nabla^2\theta=0$.
The full $U(1)$ symmetry with unrestricted gauge parameter seems to be broken but
this is solely due to the fact that the gauge has been fixed.

Similarly, one may wonder if the restricted symmetry of (\ref{hor}) is really some
fundamental property of
the theory or just a choice of gauge for the full group of four-dimensional diffeomorphism.
If this is the case, then, one should be
able to rewrite Ho\v{r}ava's theory in a fully four-dimensional covariant way. This is
the task of the present work.

\section{A Relativistic interpretation of the Ho\v{r}ava theory}

Our aim here is to write the action (\ref{hor})
in terms of four-dimensional scalars where, for the sake of covariance, both the spacetime
metric and $n^\alpha$ are dynamical \cite{wald}.
This requirement explicitly preserves the full four-dimensional group of diffeomorphisms.

A basic geometric object entering into the  action (\ref{hor}) is the extrinsic curvature
$K_{ij}$. The extrinsic curvature associated with the particular foliation chosen,
is the variation $h_{\alpha\beta}$ along the temporal direction $n^\alpha$.
The extrinsic curvature therefore encodes the
``time'' derivative of the propagating field $h_{\alpha\beta}$ and
is in fact a four-dimensional object. In a four-dimensional notation, it is written as
\be
K_{\alpha\beta}=\frac{1}{2}\pounds_n h_{\alpha\beta}\ ,
\ee
where $\pounds_n$ is the Lie-derivative along $n$.

The kinetic term of the theory can therefore be written as
\be\label{kin23}
S_{\rm kin}= \int d^4x \sqrt{-g}\left[\frac{2}{\kappa^2}\left(K_{\alpha\beta}K^{\alpha\beta}-
\lambda K^2\right)\right]\ ,
\ee
where $\lambda$ is a constant and $K=K_\alpha^\alpha$. We remark here that the action (\ref{kin23}), as it is written in terms of four-dimensional scalars, is fully diffeomorphism invariant for {\it any} $\lambda$ contrary of what claimed in \cite{hor2}.

To further define the theory we need to include ``potential'' terms which should only depend
on spatial derivatives of the metric. We realized that if $n^\alpha$ is irrotational, the space related
to $h^{\alpha\beta}$ is a sub Riemannian space-like manifold.
Therefore all curvatures computed on the foliation only depends on spatial derivatives.
Nevertheless, three dimensional curvatures (${\cal R}_{\alpha\beta\gamma\delta}$),
i.e. curvatures of the space-like foliations, are four dimensional tensors related
to the four dimensional curvatures and extrinsic curvature by the Gauss equation
\be\label{43}
{\cal R}_{\lambda\mu\nu\rho}=R_{\alpha\beta\gamma\delta}h^\alpha_\lambda
h^\beta_\mu h^\gamma_\nu h^\delta_\rho
+2 K_{\mu[\nu}K_{\rho]\lambda}\ .
\ee
${\cal R}_{\lambda\mu\nu\rho}$ contains two spatial derivatives of the spatial metric.
The covariant derivatives appearing in (\ref{hor}) may be promoted to four-dimensional
ones by using
\be
h^\alpha_\mu h^\beta_\nu h^\gamma_\delta \nabla_\alpha {\cal R}_{\beta\gamma}=D_\mu
{\cal R}_{\nu\delta}\ ,
\ee
where $D$ is the spatial covariant derivative defined by taking the four-dimensional
derivative and projecting all indices with $h$ into the spatial foliation \cite{manyfaces},
according to (\ref{dnab}). Due to the invariance
of the three dimensional space, this derivative effectively only operates on the sub-Riemannian
manifold. In addition, the
Cotton tensor (\ref{cot}) is written in a four-dimensional notation as
\be
C^{\mu\nu}=\eta^{\mu\alpha\beta}D_{\alpha}\left[{\cal R}^{^\nu}{}_\beta-\frac{1}{4}
{\cal R}\delta^\nu_\beta\right]\ ,
\ee
where $\eta^{\mu\alpha\beta}$ is the three-dimensional volume element
defined as $\eta^{\mu\alpha\beta}\equiv \eta^{\mu\alpha\beta\delta}n_{\delta}$
with the property $D_{\alpha}\eta_{\mu\nu\rho}=0$ \cite{manyfaces}.

Until now we have always assumed that $n^\alpha$ is a normalized zero vorticity vector.
To enforce that we introduce the action
\be
\label{sign}
S_{\rm norm}\sim \int d^4x \sqrt{-g}\ \left[
B^{\alpha\beta}{\cal F}_{\alpha\beta}+M^{\alpha\beta\mu\nu}B_{\alpha\beta}B_{\mu\nu}
+\rho (n^\alpha n_\alpha+1)\right]\ ,
\ee
where $\rho$, $B_{\a\b}$ and $M_{\a\b\m\n}$ are Lagrange multipliers. Varying $\r$ enforces the normalization
condition (\ref{2-7})
on $n$, whereas varying the other two multipliers enforces the Frobenius condition (\ref{2-8}) and
$B_{\a\b}=0$.
The only possible degree of freedom in $n_\alpha$ is obviously $N$ as, by using the
diffeomorphism invariance of the theory, one can always fix time to be $t=\chi$.
The fact that there is only one degree of freedom in $n_\alpha$ is clear by just noticing that the
 irrotational constraint reduce the number of degrees of freedom from four to two.\footnote{
The antisymmetric tensor ${\cal F}_{\a\b}$ obeys the four conditions
${\cal F}_{\a\b} n^\b=0$. That leaves $6-4=2$ independent components which, from the specific
form of ${\cal F}_{\a\b}$ in (\ref{2-8}), are two of the components of $n_\alpha$.} The
normalization condition finally leaves only one degree of freedom for $n_\alpha$.

To recover GR in the IR (low energies, large length scales in the adapted coordinates to $n_\alpha$),
we can now add a potential
$V({\cal R}_{\alpha\beta})$. This potential should be taken to be at most quadratic in the
three dimensional Ricci curvatures so that it does not affect the
UV (high energies, small length scales in the adapted coordinates to $n_\alpha$) behavior dominated by the term quadratic in the Cotton tensor.
In particular, by adding the term (\ref{sign}) to the action, one can extend the original
non-relativistic
detailed balance model of Ho\v{r}ava (or the simpler model of \cite{alex}) to be invariant under the full diffeomorphism group.
A four-dimensional covariant theory with
the required UV properties, compatible with the Ho\v{r}ava theory, is therefore
\begin{eqnarray}
\label{gen} S_{\rm g}&=&\int d^4 x\sqrt{-g}\ \Big\{a_0 {\cal R}+
a_1 K_{\alpha\beta}K^{\alpha\beta}+ a_2 K^2+a_3 {\cal
R}_{\alpha\beta} {\cal R}^{\alpha\beta} \cr &-&a_4
C_{\mu\nu}C^{\mu\nu} +a_5 C_{\mu\nu}{\cal R}^{\mu\nu} + a_6 {\cal
R}^2+ 2\Lambda +
B^{\alpha\beta}{\cal F}_{\alpha\beta}+M^{\alpha\beta\mu\nu}B_{\alpha\beta}B_{\mu\nu}+\cr
&+&\rho (n^\alpha n_\alpha+1)\Big\}\ ,
\end{eqnarray}
where $a_i$ are parameters. In this form of the action the independent fields are
$h^{\alpha\beta}, n^\alpha, \rho, B_{\mu\nu}$ and
$M_{\alpha\beta\gamma\delta}$.  
Note that, to define
 the spatial metric $h_{\alpha\beta}$, only the normalization condition is needed
which has been indeed introduced as a constraint for the
variational problem. The
 removal of the Frobenius conditions will only define a new theory with more degree
of freedom than the Ho\v{r}ava theory. The Frobenius conditions
define therefore the anisotropic
 theory with minimal degrees of freedom by restricting the possible solutions for $n_\alpha$.

The theory (\ref{gen}) differs from GR  mainly by the introduction of a
new vectorial degree of freedom $n^{\alpha}$ which treats undemocratically
the spaces parallel and orthogonal to it.
In any case, as already
stressed before, this new degree of freedom does not break the covariance of
the theory, i.e. the action (\ref{gen}) is invariant under the full group
of diffeomorphisms
\be\label{diff}
\delta x^\mu=\zeta^\mu(x^\alpha)\, , ~~~~\delta g_{\mu\nu}=\nabla_\mu \zeta_\nu+\nabla_\nu \zeta_\mu\, , ~~~~\delta n_\alpha=
\zeta^\mu\nabla_\mu n_\alpha+n_\mu\nabla_\alpha \zeta^\mu\, ,\label{tr}
\ee
as compared to (\ref{restr}).

We stress that
both $g_{\alpha\beta}$ and $n_{\alpha}$ in (\ref{gen}) are dynamical in compatibility
with covariance \cite{wald}.
However, by field redefining $g_{\alpha\beta}=h_{\alpha\beta}-n_\alpha n_\beta$, $n_{\alpha}$
is no longer dynamical. Therefore, $n^{\alpha}$ (and in particular $N$) is a Lagrange multiplier for the field theory of $h_{\alpha\beta}$.
Had the theory been constructed from a rotational vector, say $u^{\alpha}$, then it
would have had an extra vectorial degree of freedom.

We have finally accomplished our goal. The action (\ref{gen}) has indeed two time derivatives and higher
space derivatives in the adapted coordinate frame on $n_\alpha$, the ADM observer. However, in a
boosted  frame, higher time derivatives will be induced by space derivatives leading presumably
to acausal propagation and instabilities. Nevertheless, although locally any boosted observer
is as good as any other, in this theory it is not the case globally.
In fact, for the privileged ADM observer there is  a well posed Cauchy problem
with two initial data that gives rise to physical solutions specified by appropriate boundary
conditions at spatial infinity. Of course, the boundary conditions
at the ADM spatial infinity correspond to very special boundary conditions for a
boosted observer.\footnote{CG thank Oriol Pujolas for pointing this out to him.}
A non-ADM observer will probe solutions with incredibly fine tuned initial
conditions such that to cancel acausal propagation
and instabilities. However, the fine tuning is only apparent as the correct
boundary conditions can only be imposed globally form an ADM observer.
Thus, although the boosted ADM observer may account higher time derivatives in his effective action,
his boundary conditions (the boosted ADM ones) will kill all the unphysical
 modes related to the higher time derivative solutions.
In other words, boosting cannot turn normal fields into ghosts.

One may finally also check that all couplings of our theory at the Lifshitz point are
related to relevant operators.
Moreover, this theory can be expanded on a flat background which is a vacuum solution.
In this sense, Lorentz invariance is preserved.
The precise correspondence with Ho\v{r}ava's Lagrangian (\ref{hor}) is derived
by fixing the ADM gauge and specifying the parameters to be
\be
&&a_0=\frac{\kappa^2 \mu^2}{8(1-3\lambda)c}\Lambda_W\ , \quad
a_1=\frac{2c}{\kappa^2}\ , \quad a_2=-\frac{2c\lambda}{\kappa^2}\ ,
\quad a_3=-\frac{\kappa^2\mu^2}{8 c} \ , \quad  a_4=\frac{\kappa^2}{2 w^4 c} \nonumber \\
&&
a_5=\frac{\kappa^2 \mu}{2w^2 c}\ , \qquad\quad\qquad
a_6=\frac{\kappa^2 \mu^2}{8 c}\frac{1-4\lambda}{(1-3\lambda)}\ ,
\quad \Lambda=-\frac{3\kappa^2\mu^2}{16(1-3\lambda)c}\Lambda_W^2\ ,
\ee
in order to obtain the terms of the action arising from the detailed balance condition.
Note
that the scaling anisotropy between space and time may be
reintroduced by writing $dx_0=c dt$. To this action however a soft detailed balance
breaking term must be added in order to have the correct IR GR limit of the theory
\cite{alex,pope,Takahashi,charmousis} (for a detailed account of various possibilities
see \cite{sotiriou}) .
We will then consider the relativistic extension of the minimal theory as in \cite{alex}.

The simplest covariant Lagrangian which is diffeomorphism invariant,
power counting renormalizable at the Lifshitz fixed point and with the correct IR behaviour is then
\begin{eqnarray}
S=\int \!\!&&\!\!d^4x \sqrt{-g}\Big\{\frac{2}{\kappa^2}\left(K_{\alpha\beta}K^{\alpha\beta}-
\lambda K^2\right)-\frac{\kappa^2}{2w^4}C_{\alpha\beta}C^{\alpha\beta}+\frac{\kappa^2\mu}{2w^2}\eta^{\alpha\beta\gamma}
{\cal R}_{\alpha\epsilon}D_{\beta}{\cal R}^\epsilon{}_\gamma\cr  \label{simplest}
\!\!\!\!&\!\!-\!\!&\frac{\kappa^2\mu^2}{8}{\cal R}_{\alpha\beta}{\cal R}^{\alpha\beta}+\frac{\kappa^2\mu^2}{8(1-3\lambda)}
\frac{1-4\lambda}{4}{\cal R}^2+
\mu^4{\cal R}
+
B^{\alpha\beta}{\cal F}_{\alpha\beta}+M^{\alpha\beta\mu\nu}B_{\alpha\beta}B_{\mu\nu}+\cr
&+&\rho (n^\alpha n_\alpha+1)\Big\}\ .
\end{eqnarray}
Let us now count the propagating degrees of freedom of the graviton $h_{\alpha\beta}$.
To do that it is simpler to re-write the theory in terms of the two fields $g_{\alpha\beta}$
and $n_\alpha$. Since the theory is fully diffeomorphism invariant the tensor $g_{\alpha\beta}$
has only two propagating degrees of freedom and, as we have already stressed, $n_\alpha$
has only one degree of freedom. By conservation of the number of degrees of freedom under
field redefinition one than discover that the propagating graviton $h_{\alpha\beta}$
contains three degrees of freedom. One might now ask whether the extra degree of
freedom with respect to the usual graviton propagation in GR might go away for certain choice of parameters.
Generically this is not the case.\footnote{Note however that in specific cases the number
of propagating degrees of freedom can be lower as we shall soon discuss.}
In fact, in order to remove the extra degree of freedom in the system, the action should
not explicitly depend upon $n$. This would be only in general possible if the action was
written in terms of four-dimensional curvatures.
In the $\lambda=1$ case, the extra propagating mode should freeze at the IR as
the theory tends to GR.

Matter can now be consistently added in a diffeomorphism invariant way to the theory
(for issues related to adding matter to the Ho\v{r}ava theory see \cite{calcagni, charmousis}),
by minimally coupling it not only to the dynamical field $h_{\alpha\beta}$,
but also to the combination $g_{\alpha\beta}=h_{\alpha\beta}-n_\alpha n_\beta$. Schematically
\be
S=\int d^4x \sqrt{-g}
\left[R + L_{\rm grav}+L_{\rm m}\right]\ ,
\ee
where $L_{\rm m}$ is the  matter Lagrangian and $L_{\rm grav}=(\lambda-1)K^2-\tilde V({\cal R}_{\alpha\beta},
D_{\mu}{\cal R}_{\alpha\beta})$.

Since this theory is invariant under four-dimensional diffeomorphisms one has the following Bianchi-type identities
\be
\nabla_{\alpha}T^{\alpha\beta}_{\rm grav}=0\ , \qquad \nabla_{\alpha}T^{\alpha\beta}_{\rm m}=0\ , \label{tg}
\ee
where $T_{\rm grav,m}$ are respectively the energy momentum tensors of $L_{\rm grav}$ and $L_{\rm m}$.
The conservation of the energy momentum tensor for matter alone is equivalent to the field
equations of matter.

Let us suppose that the full set of diffeomorphisms for the gravity sector was not allowed,
as in the Ho\v{r}ava theory. The field equations of motion for the matter would
still imply the conservation of the energy
momentum tensor (if the matter Lagrangian is kept to be a four dimensional scalar).
Then, the gravity equations would require
$\nabla_{\alpha}T^{\alpha\beta}_{\rm grav}=0$ (the Einstein tensor is conserved identically).
This is equivalent to
re-introducing the full diffeomorphism group of transformations
(for a lucid explanation of this equivalence see \cite{lucid}).

A related issue here concerns the ``potentially'' strong coupling problem
reported in \cite{charmousis}.
Although in the analysis of \cite{charmousis} care has been given to restore the
full four-dimensional diffeomorphism
invariance in the gravity sector (so that $\nabla_{\alpha}T^{\alpha\beta}_{\rm grav}=0$ holds),
the matter
sector is not invariant under four-dimensional diffeomorphisms making the reported strong coupling problem of
Ho\v{r}ava
gravity questionable. In our approach, the coupling of the covariant theory (\ref{gen}) to a four-dimensional
diffeomorphism invariant matter sector is straightforward. In this respect, the issue of the
strong coupling problem discussed in \cite{charmousis} could consistently be
analyzed (see also \cite{mukohyama}).
This is however beyond the
scope of the present work.

\section{An example: cosmological perturbations}

Our point is that the Ho\v{r}ava theory should be embedded into a Tensor-Vector theory where
gravity propagates with the field $h_{\alpha\beta}$.
By explicitly using the ADM decomposition the constraint (\ref{sign})
is automatically satisfied and one recover the action of \cite{hor2}.
However, one has to now remember that this theory comes from a fully diffeomorphism
invariant action where the gauge parameter $\zeta^\mu$ in (\ref{diff}) is partially fixed
such that $\chi=t$. This has important consequences for example in cosmological perturbations
as we shall now show.

Suppose one wishes to study scalar perturbations at the linearized level.
As we have pointed out several times,
the dynamical covariant system is formed by gravity $g_{\alpha\beta}$ and
the field $n_\alpha$.
Using the gauge choice $\chi=t$, scalar perturbations in a Friedmann--Robertson--Walker background
\be
ds^2=a^2(t)\left[-dt^2+\delta_{ij}dx^i dx^j\right]\ ,
\ee
might be generically parametrized as \cite{riotto2}
\be\label{frw}
ds^2=a^2\left[(-1-2A)dt^2+2\partial_i B dt dx^i+\left((1-2\psi)\delta_{ij}+
\nabla_i\nabla_j E\right)dx^idx^j\right]\ .
\ee
If one wishes to keep the same foliation
of the background metric,
one has to fix in (\ref{diff}) $\zeta^0=\zeta^0(t)$ \cite{riotto2}. This partial gauge fixing still leaves 
the relativistic theory invariant under the smaller group of foliation preserving diffeomorphisms, as in the Ho\v{r}ava theory \cite{hor2}. 

Let us now parametrize the diffeomorphisms in the usual way \cite{riotto2}
\be
\delta t=\zeta^0(x^\mu)\ ,\quad \delta x^i=\nabla^i\beta(x^\mu)+v^i(x^\mu)\ ,\quad \partial_i v^i=0\ ,
\ee
then we have the following transformations for the scalar modes:
\begin{eqnarray}\label{gaugemetric}
A&\rightarrow& A-\dot\zeta^0-\frac{\dot a}{a}\zeta^0\ ,\qquad \cr
B&\rightarrow& B+\zeta^0+\dot \beta\ ,\qquad \cr
\psi&\rightarrow& \psi-\frac{1}{3}\nabla^2\beta+\frac{\dot a}{a}\zeta^0\ ,
\cr
E&\rightarrow& E+2\beta\ .
\end{eqnarray}
Fixing $\zeta^0$ to preserve the foliation, leaves only one degree of
 freedom ($\beta$) in (\ref{gaugemetric}) in contradiction with GR
in which the degrees of freedom are two ($\zeta^0,\beta$).
One can then easily see that three of the graviton components cannot be set to zero.
Thus, in principle gravity generically contains three modes.
Whether the three modes are all propagating or not depends really
on the background (and the perturbative level).
In fact in \cite{riotto}, in presence of matter, only one propagating linear mode has been found. Also in GR a similar thing happens. There, on a cosmological background, the physical modes are two but only one of them, or none in absence of a scalar source, propagates.

\section{Conclusions}

Theories that extend GR with additional vectorial degrees of freedom,
but nevertheless invariant under the full group of diffeomorphisms, are not new. For example, the
Bekenstein TeVeS theory of gravity \cite{bekenstein} and the Einstein--Aether theory \cite{ether}
or the ``preferred frame'' gravity theories of
Jacobson and Mattingly \cite{preferred} are of
this kind. Related to our present context, one may also write down a covariant theory
which couples non-minimally to a specific (dynamical) perfect fluid stress-tensor with curvatures.
In this case, one might obtain a higher spatial derivative theory for the graviton \cite{odin}.
This theory however introduces many extra degrees of freedom into the
purely gravitational action and it is also not clear whether the GR limit at IR is obtained.

In this paper we showed that the Ho\v{r}ava theory for a power counting renormalizable quantum
gravity, tending to GR at the IR,
is nothing else but a gauge fixed diffeomorphism invariant Tensor-Vector theory.
Indeed, the full diffeomorphism invariance of the Ho\v{r}ava theory can be restored by
introducing a Lagrange multiplier which normalizes the irrotational time-like vector field $n^\alpha$
defining a space-like foliation of the spacetime manifold. This, seems also to resolve potential problems related to the canonical formulation of the Ho\v{r}ava theory \cite{yi}.

As seen from the four-dimensional metric ($g_{\alpha\beta}$) perspective,
the theory generically propagates two degrees of freedom from the metric, as it should happen for a
diffeomorphism invariant theory of gravity, and one degree of freedom from $n_\alpha$.
However, this theory can also be studied by using the fields
$h_{\alpha\beta}(=g_{\alpha\beta}+n_\alpha n_\beta)$ and $n_\alpha$.
In this base only $h_{\alpha\beta}$ propagates encoding all degrees of freedom
and results in a power counting renormalizable field theory.
Its orthogonal projection to $n_\alpha$, defines an invariant three-dimensional
Riemannian sub-manifold parametrized by the time-like direction $n^\alpha$.
Therefore, $n^\alpha$ only acts as a ``Lagrangian multiplier''
for the true degrees of freedom in $h_{\alpha\beta}$.

Fixing the spacetime metric to be in the ADM form as in \cite{hor2}
partially fixes the gauge freedom of the theory by requiring $\chi=t$,
 where $n_\alpha\equiv -N\partial_\alpha\chi$,
restricting the diffeomorphisms invariance of the theory to be only foliation preserving.

Although the theory we presented is fully diffeomorphism invariant,
it treats anisotropically the space-like and the time-like sub-spaces
delineated by the local light cones of $n$.
This might seems to violate Lorentz invariance locally.
However, as the vacuum solution of the theory is still Minkowski,
the linearized graviton Lagrangian is still invariant under boosts,
although the graviton dispersion relations are different from the
usual GR one at UV scales.

Let us now instead start from the Ho\v{r}ava theory where $n$ is an external field
and where only the restricted group of diffeomorphisms (\ref{restr}) is allowed.
In this case if matter is taken to be standard, i.e. it is described by a
four dimensional scalar Lagrangian,
general covariance
would imply that the divergence of the total energy-momentum tensor of all
geometrical quantities should vanishes when the matter field equations are satisfied.
This is a very strict physical constraint necessary for the consistency
of a (diffeomorphisms invariant) matter-gravity system.
If a preferred frame for the gravity sector is then introduced and general
covariance is abandoned, the conservation of the matter energy momentum tensor
would nevertheless reintroduce the full diffeomorphism group as a constraint equation for gravity.
Of course this would not be the case if the matter Lagrangian is not described by a four dimensional scalar.

The above are reminiscent of a case described in \cite{LoVio, Jackiw}.
In this work a modified gravity of the form
\be
S=\frac{1}{2\kappa^2}\int d^4 x \sqrt{-g}\left(R-\frac{1}{2}u_\mu K^\mu\right)\ ,
\ee
was considered, where the vector $u^\mu=(\frac{1}{m},0,0,0)$ is a fixed background
vector and $K^\mu$ is the gravitational Chern--Simon current whose divergence is
 the Chern--Pontryagin density ($\partial_\mu K^\mu=\frac{1}{2}{}^*R\ R$). The existence
 of the fixed vector $u^\mu$ apparently violates four-dimensional diffeomorphism invariance and Lorentz
 invariance. However, as it has been shown in \cite{Jackiw}, the theory is diffeomorphism
invariant and there are no Lorentz violating effects. In fact, one may start with a
diffeomorphism invariant theory with a varying $u^\mu$. Then  diffeomorphism invariance
 may be used to fix a particular gauge for $u^\mu$. This choice seems to restrict
 the diffeomorphism group but this is because we consume partially the diffeomorphism group to fix
a particular frame.

Thus, in general, there are cases
 where what appears to be a symmetry violation is just due to a gauge choice. In the
Ho\v{r}ava theory the apparent violation of four-dimensional covariance is just
because it is written in a specific gauge, specified by the ADM frame. In turn, this can be
used because of the four-dimensional covariance of the theory (and the corresponding constraints).
Thus, the Ho\v{r}ava's theory coupled to standard matter is a particular case of the general class
of a relativistic, fully
covariant action (\ref{gen}). If this were not the case,
then there would be problems with the conservation of the energy-momentum tensor for matter covariant
Lagrangians. The reason is that, apart form potential inconsistencies, violation of covariance in the
gravity sector would be
transmitted to the matter sector by graviton loops leading to ill defined matter theories.
We conclude by stressing that our covariant theory allows
consistent coupling of gravity with matter in contrast with the original
Ho\v{r}ava theory \cite{hor2}.

\section{Acknowledgements}
CG wishes to thank Roberto Percacci for comments on Quantum Gravity fixed points, Donato Bini,
Jose Martin-Garcia, Eric Gourgoulhon for discussions on the general 3+1 and
ADM formalisms, Yi Pang for useful comments on the variational problem for the relativistic action
and finally Antonio Riotto for discussions on the gauge transformations.
CG thanks CERN for hospitality during the preparation of this work.
AK wishes to thank support from the PEVE-NTUA-2007/10079 program. This work is
partially supported by the European Research and Training Network MRTPN-CT-2006.
\section{Erratum}

The definition (\ref{43}) is valid only if the Frobenius conditions (\ref{2-8}) are satisfied. 
This is due to the fact that 
the Frobenius conditions are non-holonomic constraints. In this case, the definition of 
$n_\alpha$ in (\ref{def}) together 
with the normalization (\ref{2-7}), i.e.,
\be
n_\alpha=-\frac{\partial_\alpha \chi}{\sqrt{-\partial_\beta \chi\partial^\beta \chi}}\, , \label{n}
\ee
 must be used in the action (\ref{gen}). The  constraints (\ref{sign}) are then 
automatically satisfied
and the theory so written  is explicitly a tensor-scalar theory, where the scalar $\chi$
represents the degree of 
freedom  of $n_\alpha$.  

In summary, once (\ref{n}) is directly used in (\ref{gen}),
$g^{\alpha\beta}$ and $\chi$ become the independent fields of the variational problem for the theory (\ref{gen}), in compatibility with \cite{pujolas}.

\end{document}